\documentclass[%
reprint,
superscriptaddress,
amsmath,amssymb,amsthm,mathrsfs,
aps,
prb,
]{revtex4-2}

\usepackage{graphicx}
\usepackage{dcolumn}
\usepackage{bm}
\usepackage[usenames]{color}
\usepackage{soul}
\usepackage[ 
margin=0.75in,
]{geometry}
\usepackage[normalem]{ulem}
\usepackage[T1]{fontenc}

\usepackage{ragged2e} 
\newcounter{figctr}
\renewcommand{\thefigctr}{\arabic{figctr}}
\newcommand{\figcaption}[1]{
  \refstepcounter{figctr}
    \begin{minipage}{\linewidth}
      \noindent\justifying \textbf{Fig.~\thefigctr:}~#1
    \end{minipage}
}

\graphicspath{{./Figures/}}
\def\TN{$T_\mathrm{N}$}

\def\muB{$\mu_{\mathrm{B}}$}

\begin{document}
	
\title{
	Tuning the magnetic properties of the spin-split antiferromagnet MnTe through pressure
}
	
\author{Edison P. Carlisle}
\affiliation{ %
    Department of Physics and Astronomy, Brigham Young University, Provo, Utah 84602, USA.}

\author{George Yumnam}
\affiliation{Materials Science and Technology Division, Oak Ridge National Laboratory, Oak Ridge, Tennessee 37830, USA}

\author{Stuart Calder}
\affiliation{Neutron Scattering Division, Oak Ridge National Laboratory, Oak Ridge, Tennessee 37831, USA}

\author{Bianca Haberl}
\affiliation{Neutron Scattering Division, Oak Ridge National Laboratory, Oak Ridge, Tennessee 37831, USA}

\author{Jia-Xin Xiong}
\affiliation{Renewable and Sustainable Energy Institute, University of Colorado Boulder, Boulder, Colorado 80309, USA}

\author{Michael A. McGuire}
\affiliation{Materials Science and Technology Division, Oak Ridge National Laboratory, Oak Ridge, Tennessee 37830, USA}

\author{Alex Zunger}
\affiliation{Renewable and Sustainable Energy Institute, University of Colorado Boulder, Boulder, Colorado 80309, USA}

\author{Rapha\"el P. Hermann}
\affiliation{Materials Science and Technology Division, Oak Ridge National Laboratory, Oak Ridge, Tennessee 37830, USA}

\author{Benjamin A. Frandsen}
\affiliation{ %
    Department of Physics and Astronomy, Brigham Young University, Provo, Utah 84602, USA.}
\email{benfrandsen@byu.edu}

\begin{abstract}
The hexagonal antiferromagnet MnTe has attracted enormous interest as a prototypical example of a spin-compensated magnet in which the combination of crystal and spin symmetries lifts the spin degeneracy of the electron bands without the need for spin-orbit coupling, a phenomenon called non-relativistic spin splitting (NRSS). Subgroups of NRSS are determined by the specific spin-interconverting symmetry that connects the two opposite-spin sublattices. In MnTe, this symmetry is rotation, leading to the subgroup with spin splitting away from the Brillouin zone center, often called altermagnetism. MnTe also has the largest spontaneous magnetovolume effect of any known antiferromagnet, implying strong coupling between the magnetic moment and volume. This magnetostructural coupling offers a potential knob for tuning the spin-splitting properties of MnTe. Here, we use neutron diffraction with \textit{in situ} applied pressure to determine the effects of pressure on the magnetic properties of MnTe and further explore this magnetostructural coupling. We find that applying pressure significantly increases the N\'eel temperature but decreases the ordered magnetic moment. We explain this as a consequence of strengthened magnetic exchange interactions under pressure, resulting in higher \TN, with a simultaneous reduction of the local moment of individual Mn atoms, described here via density functional theory. This reflects the increased orbital hybridization and electron delocalization with pressure. These results shed light on the competition between magnetic exchange interactions and the strength of individual magnetic moments and show that the magnetic properties of MnTe can be controlled by pressure, opening the door to improved properties for spintronic applications through tuning via physical or chemical pressure.\footnote{This manuscript has been authored by UT-Battelle, LLC, under contract DE-AC05-00OR22725 with the US Department of Energy (DOE). The US government retains and the publisher, by accepting the article for publication, acknowledges that the US government retains a nonexclusive, paid-up, irrevocable, worldwide license to publish or reproduce the published form of this manuscript, or allow others to do so, for US government purposes. DOE will provide public access to these results of federally sponsored research in accordance with the DOE Public Access Plan (\href{https://www.energy.gov/doe-public-access-plan}{https://www.energy.gov/doe-public-access-plan}).}
\end{abstract}
	
\maketitle

\section{Introduction}
In recent years, hexagonal manganese telluride (MnTe) has emerged as an important material in a variety of contexts spanning both fundamental research and technological application. An antiferromagnet (AFM) with a N\'eel temperature of 307~K~\cite{walther1967}, MnTe~\cite{gonzalez2023spontaneous,mazin2023altermagnetism,lee2024broken,krempasky2024altermagnetic}, along with other AFM compounds (MnF$_2$~\cite{yuan2020giant}, orthorhombic LaMnO$_3$~\cite{yuan2021prediction}, rhombohedral MnTiO$_3$~\cite{yuan2021prediction}), has been highlighted as an example of the newly recognized class of collinear, spin-compensated magnets that break the spin degeneracy of the electronic band structure without the need for spin-orbit coupling (SOC), termed non-relativistic spin splitting (NRSS)~\cite{yuan2020giant,yuan2021prediction}. In nonmagnetic materials, spin degeneracy can be lifted by broken inversion symmetry and SOC, as in the Rashba/Dresselhaus effects. In contrast, NRSS is a consequence of specific enabling ``broken symmetries'' in AFMs: Whereas ordinary (``symmetry unbroken'') N\'eel AFMs such as NiO and CuMnAs are spin compensated and, in the absence of SOC, have degenerate spin bands, symmetry-broken AFMs violate space- plus time-reversal symmetry and translational- plus spin-rotation. The lifting of the spin degeneracy of the energy bands (``spin splitting'') occurs at certain locations in the Brillouin zone (BZ) depending on the additional auxiliary spin-interconverting symmetry conditions that connect the two opposite-spin sublattices. The auxiliary spin-interconverting symmetries of any given AFM compound can be identified from the crystallographic space group plus the direction of the magnetic moments of the magnetic ions.  If this auxiliary symmetry is rotation (with optional reflection), the ensuing NRSS rotational subgroup (often dubbed ``altermagnetism'') occurs away from the BZ center along low-symmetry directions, as illustrated in the present work for MnTe. Other auxiliary symmetries create other subgroups of spin splitting, such as NRSS at the BZ center~\cite{yuan;prl24}. Due to their unique combination of favorable characteristics of both FM materials (e.g. spin-split bands) and traditional AFM materials (e.g. lack of stray magnetic fields, ultrafast switching), NRSS AFMs have stimulated enormous interest for their potential applications in spintronics, topological matter, thermoelectrics, and more~\cite{vsmejkal2022beyond,vsmejkal2022emerging}.


Spin splitting of the Mn 3$d$ bands in MnTe has been predicted theoretically through band-structure calculations and verified experimentally through direct observation of the spin splitting~\cite{krempasky2024altermagnetic}, as well as through observation of the spontaneous anomalous Hall effect~\cite{gonzalez2023spontaneous,kluczyk2024coexistence}. Other aspects of MnTe unrelated to altermagnetism have also drawn attention for potential spintronics applications, including manipulation of the magnetic domain structure~\cite{kriegner2016multiple}, ultrafast spin-charge coupling~\cite{bossini2021femtosecond}, and tuning of the magnetic anisotropy and magnon gap by doping~\cite{mosely2022giant,yumnam_prb24}. Beyond that, MnTe has attracted research interest as the magnetic component of the intrinsic magnetic topological insulator MnBi$_2$Te$_4$~\cite{deng2020quantum} and as a high-performance thermoelectric~\cite{xu2017performance,ren2017synergistic,dong2018lead,deng2021high,xiong2023lattice,zulkifal2023multiple}, where short-range magnetic correlations well above \TN\ significantly boost the thermopower~\cite{zheng2019paramagnon,polash2021understanding,baral2022real}.

Here, we focus on the remarkable magnetostructural properties exhibited by MnTe. A recent study~\cite{baral2023} has shown that MnTe possesses a giant spontaneous magnetovolume effect with a volume contraction of nearly 1\% across \TN, the largest of any known AFM outside of systems with a collapse of the magnetic moment such as YMn$_2$~\cite{nakamura1983magnetovolume}. The volume contraction scales linearly with the magnitude of the local magnetic correlations, rather than the usual quadratic scaling. This scaling arises due to a novel trilinear magnetostructural coupling between lattice strain and superposed short-range and longer-range antiferromagnetic domains of the magnetic order parameter~\cite{baral2023}. The pronounced lattice contraction in response to the magnetic order suggests that the magnetic order may in turn respond to reduction of the unit cell volume via applied pressure.

Early measurements of the temperature-dependent resistivity of MnTe at different applied pressures did, in fact, suggest that the N\'eel temperature increases with pressure~\cite{ozawa1966effect}, supporting the viability of pressure as a means of tuning the magnetic properties of MnTe. However, no direct studies of the magnetic order in response to pressure are found in the experimental literature. We therefore aim to study the effect that pressure has on the magnetism of MnTe, including both the ordering temperature and the magnitude of the magnetic order parameter. Indeed, how the ordering temperature and magnetic moment respond individually to pressure in any given material is an important question for the fundamental materials physics of magnetostructural coupling. Do ordering temperature and ordered moment respond uniformly to pressure (e.g. both are enhanced or both are suppressed by pressure), or can they exhibit opposite responses (e.g. an increase in ordering temperature with a decrease in magnetic moment or vice versa)? Beyond the fundamental interest, explorations of magnetostructural coupling and pressure-tuned magnetism also hold practical significance for potential applications by, for example, raising the magnetic ordering temperature via physical pressure or chemical pressure.

In this paper, we elucidate the effects of applied pressure on the magnetic moment and N\'eel temperature of MnTe via neutron diffraction with \textit{in situ} hydrostatic pressure. We observe that with increasing pressure, the N\'eel temperature increases while the ordered magnetic moment tends to decrease. This unusual combination of seemingly opposite tendencies of the magnetic phase transition results from the strengthening of the magnetic exchange interactions during compression of the unit cell volume, thus raising \TN, and the simultaneous reduction of the local magnetic moment on the Mn site due to increased orbital hybridization and electron delocalization, thus decreasing the ordered magnetic moment. We support the experimental results with density functional theory (DFT) calculations to confirm the effects of pressure on the magnetic moment, which also reveal an enhancement of the NRSS in the valence band and a reduction in the conduction band. These results establish pressure as a vehicle for tuning NRSS properties and highlight the competition between magnetic exchange interactions and the magnetic strength of isolated moments.

\section{Methods}

\subsection{Neutron diffraction procedure and analysis}
The polycrystalline MnTe sample was made by reacting the elements in an alumina crucible sealed inside an evacuated silica ampoule as described in Ref.~\cite{mosely2022giant}. The sample was stable in air; no additional impurity was observed after one month of air exposure.

Neutron powder diffraction was performed at Oak Ridge National Laboratory's HB-2A POWDER beamline with an incident wavelength of 2.41~\AA. The sample was loaded into a Teflon container which in turn was loaded into a piston cylinder cell consisting of a main Al cylinder with a prestressed CuBe insert as well as a WC piston assembly with optical access for fluoresence measurements~\cite{podlesnyak2018clamp}. Fluorinert was used as pressure medium as a prior test revealed minimal evaporation under the temperature conditions used. Further, fluorinert remains hydrostatic under the pressure conditions used here~\cite{varga2003fluorinert}. This cell was chosen to minimize background contamination at MnTe's magnetic Bragg peaks. Neutron diffraction patterns were collected at applied pressures of 0, 0.3, 0.5, and 1~GPa, where we used ruby fluorescence and its associated pressure equation of state~\cite{dewaele2008compression} to measure these pressures. Using the equation of state found in Ref.~\cite{mimasaka1987pressure} and lattice parameters from our Rietveld refinements, we find that our pressures are accurate within 0.1~GPa. For each pressure, we used the cryostat at HB-2A to acquire the full powder diffraction pattern at 250 and 350~K. We also acquired full powder diffraction patterns for at least one pressure at $T=$280, 295, 320, and 370~K. In these diffraction patterns the reflections 001, 100, 002 and 101 are well resolved and separated from any pressure cell scattering. However, most other peaks were heavily contaminated by the CuBe pressure cell background, so we had a limited $2\theta$ range $(17^{\circ}<2\theta<~55^{\circ})$ to do Rietveld refinements.

Rietveld refinements were performed using FullProf~\cite{fullprof}. We started by fitting lattice parameters and a preferred orientation parameter to each of the patterns collected at 350~K. Because $T_N<350$~K for all pressures, the observed peaks from MnTe are purely nuclear with no magnetic contribution. The preferred orientation parameter $G_1$ was nearly identical for fits. Then, for all other temperatures, we fixed the preferred orientation parameter for each pressure to the values found at 350~K. $B_\mathrm{iso}$ was fixed to 1.0~\AA$^2$ for all fits, since the limited usable $2\theta$ range imposed by the pressure cell precluded reliable refinements of this parameter. The lattice parameters and ordered moment were then refined for all lower-temperature data sets.

We also conducted an order parameter (OP) scan at each pressure. The OP scans measured the intensity of MnTe's antiferromagnetic 001 peak using a single detector of HB-2A's detector bank. Because we only needed data collected at a single $2\theta$, we did not need to rotate the detector to cover more angles as is required for a normal diffraction pattern at HB-2A. The OP scans allow us to precisely determine the N\'eel temperature (\TN) at each applied pressure, because above \TN\ the magnetic contribution to the Bragg peaks vanishes. For 0~GPa, we scanned from 250~K to 350~K in steps of 2~K, for 1~GPa we scanned from 270~K to 370~K in steps of 1~K, and for the other pressures we scanned from 270~K to 350~K in steps of 1~K.

\subsection{DFT calculations of electronic properties}
We used a plane-wave pseudopotential density functional theory (DFT) method to perform electronic calculations as implemented by the Vienna Ab initio Simulation Package (VASP)~\cite{kresse1996efficiency,kresse1996efficient}. We utilized the projector augmented-wave (PAW) method~\cite{blochl1994projector,kresse1999ultrasoft} with the Perdew-Burke-Ernzerhof (PBE) exchange-correlation functional within the generalized gradient approximation (GGA)~\cite{perdew1996generalized} plus an on-site potential term $U$ of 4~eV taken from Ref.~\cite{alaei2025} to correct the spurious self-interaction errors in Mn-$d$ orbitals. We emphasize that this on-site potential term $U$ is different from the Mott-Hubbard $U$ that captures strong correlation effects from the interelectronic interactions, and that the on-site $U$ should not normally change when performing the DFT calculations with different values of the lattice constants. The PAW-PBE pseudopotentials are adopted with the valence electrons in MnTe including 13 electrons in Mn (2p$^6$4s$^2$3d$^5$) and 6 electrons in Te (5s$^2$5p$^4$). We set the following parameters in our DFT calculations: (i) The cut-off total energy for the plane-wave basis was 360~eV; (ii) The convergence tolerance for total energy was $10^{-6}$~eV; (iii) The convergence tolerance of relaxation was $10^{-2}$~eV/\AA; (iv) The $\Gamma$-centered $k$-point sampling grid was $5\times5\times3$ for relaxation and static self-consistent calculation.

\section{Results}

\subsection{Neutron Powder Diffraction}

\begin{figure}
        \centering
        \centerline{\includegraphics[width=0.5\textwidth]{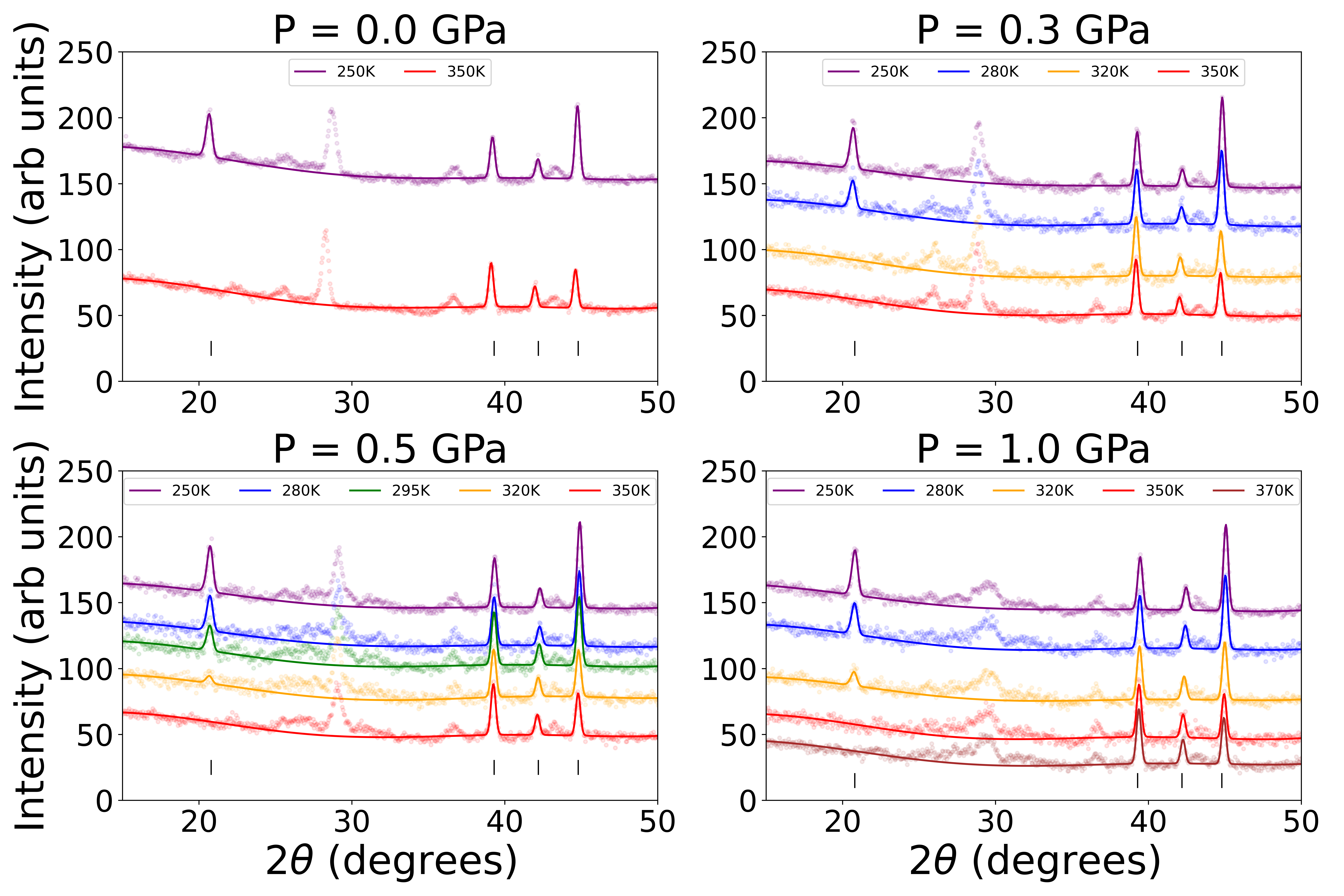}}
        \figcaption{\textbf{Neutron diffraction patterns at various temperatures and applied pressures.} The evolution of the magnetic Bragg peaks (at $\sim20^{\circ}$ and $\sim45^{\circ}$) with pressure and temperature confirm the sensitivity of the magnetic order to applied pressure. For example, we see the magnetic peaks persist up to higher temperatures as more pressure is applied. Circular symbols represent the data, and the solid curves are fits obtained via Rietveld refinement. Pressures and temperatures are indicated by the plot titles and legends. Diffraction patterns are offset vertically for clarity. Note that several peaks in the pattern are from the Teflon sample container within the pressure cell environment, which we did not attempt to include in the Rietveld refinements; only the fitted peaks are attributed to MnTe.}
        \label{fig:diffraction}
    \end{figure}

     Representative neutron powder diffraction patterns collected at various temperatures with applied pressures of 0, 0.3, 0.5, and 1~GPa, along with fits obtained via Rietveld refinement, are shown in Fig.~\ref{fig:diffraction}.
    Consistent with the known magnetic structure of bulk MnTe~\cite{kunitomi1964neutron} illustrated in Fig.~1 of the Supplementary Information~\cite{suppInfo}, the 001 magnetic Bragg peak is visible around 20$^{\circ}$ at low temperature, and another magnetic Bragg peak is co-located with the nuclear peak seen around 45$^{\circ}$. These magnetic peaks vanish for $T > T_{\mathrm{N}}$, as expected. Notably, the 001 magnetic peak is absent at 320~K for 0.3~GPa but observed at that temperature for 0.5 and 1~GPa. This confirms the increase of \TN\ with pressure, as will be further discussed below. There was no observed change in magnetic or crystal symmetry at any pressure measured. From the Rietveld refinements, we extracted the lattice parameters and ordered magnetic moment to investigate their dependence on pressure and temperature (see Table~\ref{tab:FullProf}).
    \begin{table}
        \centering
        \caption{Results of Rietveld refinements. Here, $a$ and $c$ are lattice parameters and $m$ is the ordered magnetic moment. The errors for all $c$ values are $\sim$0.001~\AA, while the errors for all $a$ values are $\lesssim 5\times10^{-4}$~\AA.}
        \begin{tabular}{ll|ccc} \hline \hline
            $P$ (GPa) & $T$ (K) & $a$ (\AA) & $c$ (\AA) & $m$ ($\mu_B$) \\
            \hline
            0 & 250 & 4.139 & 6.685 & 3.12 $\pm$ 0.08 \\
            0 & 350 & 4.148 & 6.716  & -  \\ \hline
            0.3 & 250 & 4.132 & 6.677 & 3.26 $\pm$ 0.08 \\
            0.3 & 280 & 4.135 & 6.685 & 2.66 $\pm$ 0.10 \\
            0.3 & 320 & 4.138 & 6.698 & - \\
            0.3 & 350 & 4.139 & 6.706 & - \\ \hline
            0.5 & 250 & 4.126 & 6.666 & 3.21 $\pm$ 0.08 \\
            0.5 & 280 & 4.128 & 6.669 & 2.83 $\pm$ 0.10 \\
            0.5 & 295 & 4.129 & 6.672 & 2.30 $\pm$ 0.10 \\
            0.5 & 320 & 4.131 & 6.683 & 1.30 $\pm$ 0.15 \\
            0.5 & 350 & 4.133 & 6.691 & - \\ \hline
            1 & 250 & 4.112 & 6.640 & 2.93 $\pm$ 0.07 \\
            1 & 280 & 4.114 & 6.648 & 2.43 $\pm$ 0.09 \\         
            1 & 320 & 4.116 & 6.660 & 1.61 $\pm$ 0.09 \\
            1 & 350 & 4.120 & 6.670 & - \\
            1 & 370 & 4.121 & 6.671 & - \\
            \hline
        \end{tabular}
        \label{tab:FullProf}
    \end{table}
    The lattice parameters show the expected decrease with pressure, but more complicated behavior is seen for the ordered moment. For example, when the temperature is fixed at 250~K, the ordered moment initially increases slightly with increasing pressure, then decreases for pressures beyond 0.3~GPa. As will be shown subsequently, this unusual non-monotonic change in the ordered moment as a function of pressure with fixed temperature becomes more straightforward when examining it along side the pressure-dependent N\'eel temperature.

\subsection{Increase of \TN\ With Pressure}

\begin{figure}
        \centerline{\includegraphics[width=0.5\textwidth]{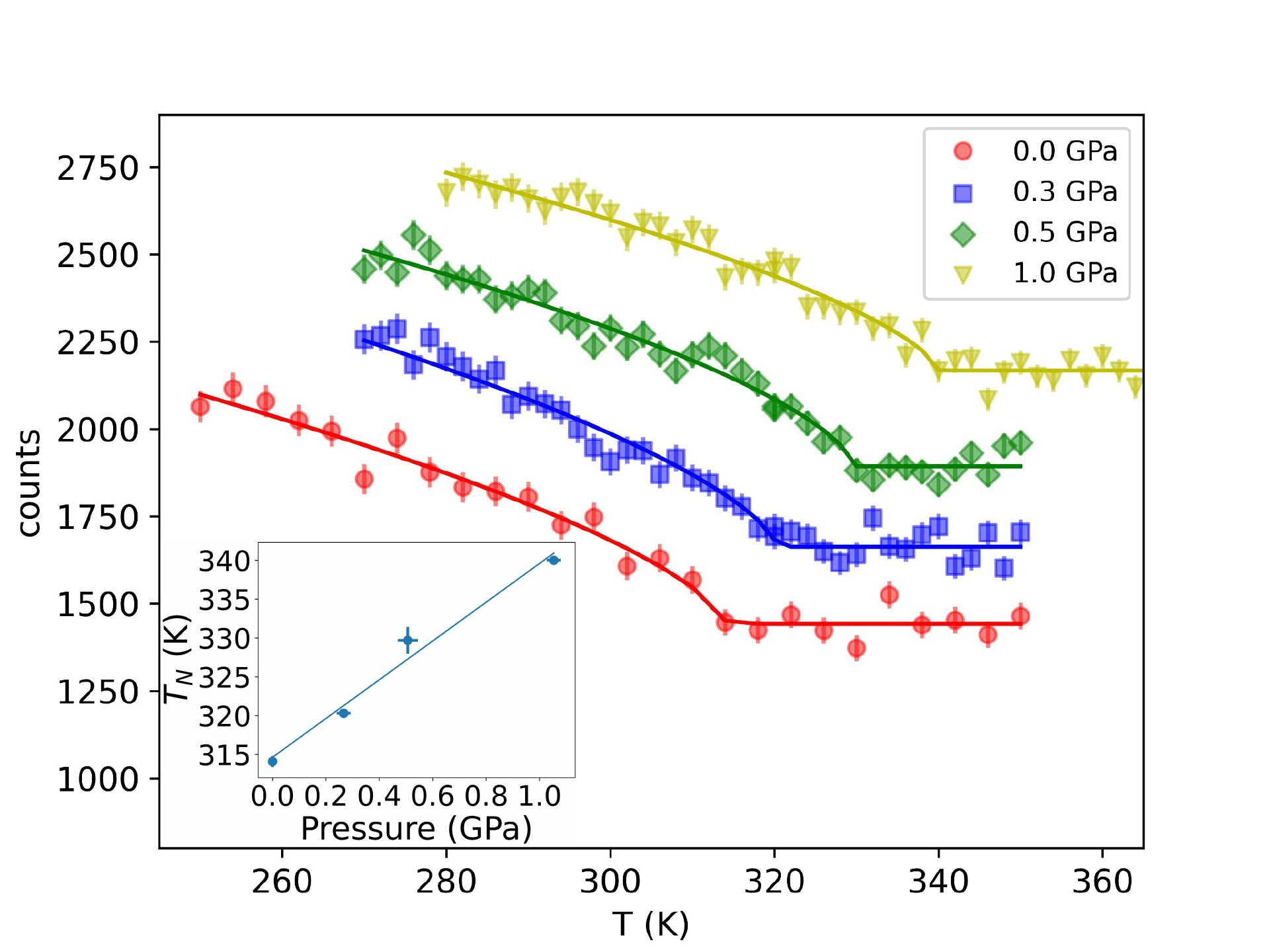}}    
        \figcaption{\textbf{Increase of \TN\ with increasing pressure.} The colored symbols are the experimental OP data as described in the main text, and the solid curves are power-law fits given by Eq.~\ref{eq:OP}. Data for $P>0$~GPa are offset vertically by a constant number of counts. Inset: Dependence of \TN\ on pressure. We note that the value of \TN\ at ambient pressure obtained from the power-law fit (314~K) is higher than the established value of 307~K, which we attribute to fitting error due to the relatively large spacing between temperature points in the ambient pressure OP scan. The OP scans at higher pressure used a much smaller temperature spacing, minimizing the potential for error in \TN.}
        \label{fig:OP}
    \end{figure}

    To find \TN\ at each pressure, we measured the intensity of the 001 magnetic Bragg peak over a fine temperature grid. We refer to this as an order parameter (OP) scan. For each OP scan, we fit the observed intensity $I$ as 
    \begin{equation}
        I = A*(1-T/T_N)^{2\beta},
        \label{eq:OP}
    \end{equation}
    where $A$ is a constant, \TN\ is the N\'eel temperature, and $\beta$ is the critical exponent. The results are shown in Fig.~\ref{fig:OP}, where we see a clear increase in \TN\ from about 315~K at ambient pressure to 340~K at 1~GPa.
    The change in \TN\ is approximately linear in pressure, with a rate of 25$\pm$0.6~K/GPa. All critical exponents are roughly $\beta=0.33$, which agrees with previous experiments~\cite{mosely2022giant, yumnam_prb24} and matches the $\beta = 0.3264$ expected from the 3D Ising model~\cite{sethna2021statistical}.

\subsection{Influence of Pressure on the Ordered Magnetic Moment}

\begin{figure}
        \centerline{\includegraphics[width=0.5\textwidth]{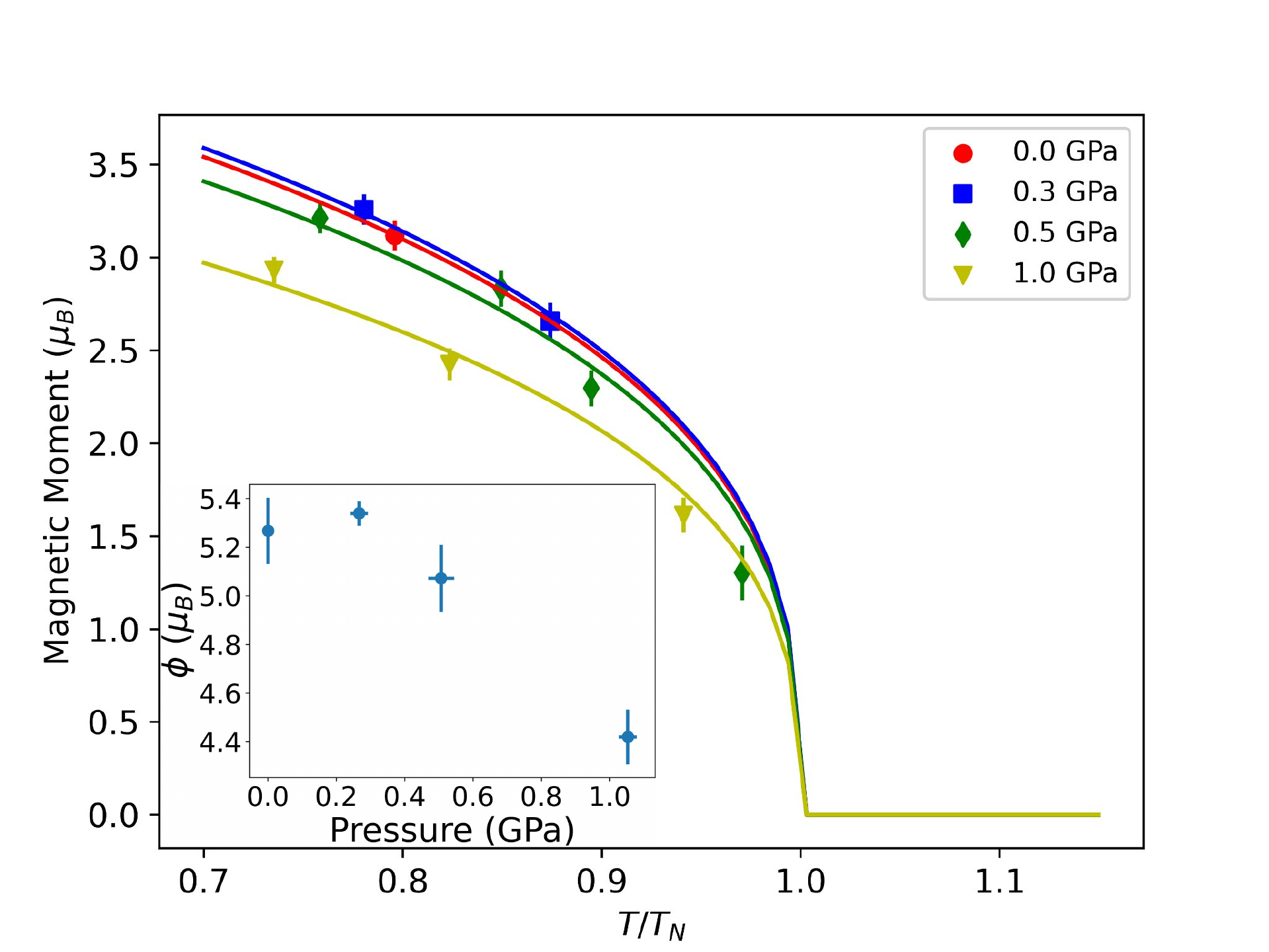}}    
        \figcaption{\textbf{Reduction of the ordered magnetic moment with increasing pressure.} The symbols represent the ordered moment plotted as a function of reduced temperature $T/T_{\mathrm{N}}$. Data at different pressures are specified by different colored symbols, as denoted in the legend. The solid curves are power-law fits obtained via Eq.~\ref{eq:magmom}. The colors of the fits match the colors of the corresponding data. Inset: Dependence of the magnetic moment scale factor $\phi$ on pressure.}
        \label{fig:MagneticMoments}
    \end{figure}

    A more complete picture of the influence of pressure on the magnetic state in MnTe emerges when we combine the information we have gleaned about both \TN\ and the ordered moment. In Fig.~\ref{fig:MagneticMoments}, we plot the ordered moment as a function of reduced temperature $T/T_{\mathrm{N}}$ for each pressure, rather than as a function of absolute temperature $T$.
    By doing so, the unusual non-monotonic relationship between the ordered moment and pressure at a fixed absolute temperature gives way to more systematic behavior: Increasing the pressure decreases the ordered moment monotonically for a given reduced temperature $T/T_{\mathrm{N}}$. For each pressure, the ordered moment near \TN\ can be expressed as a function of the reduced temperature as 
    \begin{equation}
        m(T/T_{\mathrm{N}}) = \phi * (1-T/T_{\mathrm{N}})^{\beta},
        \label{eq:magmom}
    \end{equation}
    where $m(T/T_{\mathrm{N}})$ is the ordered moment, $\phi$ is a scaling factor, and $\beta=0.33$. We fit $\phi$ to our ordered moments at each pressure. Results are shown in the inset of Fig.~\ref{fig:MagneticMoments}, where we see that $\phi$ decreases with pressure, confirming the tendency of pressure to reduce the magnitude of the ordered moment despite increasing \TN.

\section{Discussion}

The observation that increased pressure simultaneously raises \TN\ and reduces the ordered moment, two seemingly counterintuitive trends driven by the same external stimulus, stands in contrast to the more typical behavior of \TN\ and the ordered moment responding to pressure in the same direction~\cite{umebayashi1968neutron, onodera1994neutron, kernavanois2005neutron}.
However, the behavior seen in MnTe is not unprecedented---a similar effect is seen in the perovskite LaCrO$_3$ \cite{zhou2011magnetic}, for example---though detailed discussion of this phenomenon appears to be rather sparse in the literature. Here, we offer a simple conceptual model to explain the behavior observed in MnTe. 
Applying pressure to MnTe reduces the interatomic distances, which correspondingly strengthens the magnetic exchange interactions. This would tend to increase the magnetic ordering temperature. At the same time, the reduction of interatomic distances also enhances orbital hybridization, effectively making MnTe more metallic by increasing the hopping amplitude between neighboring sites. This has been confirmed experimentally by pressure-dependent resistivity measurements showing a pronounced drop in resistance with pressure~\cite{mimasaka1987pressure, wang;ChemMater22}. The greater delocalization of valence electrons in turn reduces the bare local moment on the Mn sites from the maximum allowed value corresponding to $S=5/2$. This reduction in the ordered moment would tend to reduce \TN, in competition with the strengthening of the exchange interactions. The experimentally observed increase in \TN\ indicates that the effects of the strengthened exchange interactions outweigh those of the reduced local moment.


To check the viability of this model, we can use previously estimated changes in the exchange interaction with distance~\cite{baral2023,szusz_prb06} and estimate how much the local moment must decrease to remain consistent with an increase in \TN\ of $\sim$30~K, as observed experimentally with 1~GPa of applied pressure. The ground state energy of a single site can be expressed as a function of pressure $P$ as
\begin{equation}
    E(P) = -\sum_{i=\{1,3\}} z_i\left(J_i+\frac{dJ}{dr}\Delta r_i(P)\right)\langle\mathbf{S}_0\cdot \mathbf{S}_i\rangle(P)
\end{equation}
where we sum over the first and third nearest neighbors, $z_i$ is the number of $i$th nearest neighbors, $dJ/dr$ is the change of the exchange interaction with respect to spin-separation distance, $\Delta r_i(P)$ is the pressure-induced change of distance to the $i$th nearest neighbors, and $\langle\mathbf{S}_0\cdot \mathbf{S}_i\rangle(P)$ is the product of the central spin $\mathbf{S}_0$ and the $i$th neighboring spins averaged over the $i$th coordination shell (with the pressure dependence explicitly indicated). 
Making the mean-field assumption that $E(P)$ is proportional to $T_N(P)$, we can evaluate $E(P=0)/T_N(P=0)$ using the results of the Rietveld refinements, the exchange interactions from Ref.~\cite{mu2019phonons}, and $dJ/dr$ from Ref.~\cite{baral2023}, and then solve for the required value of $\langle\mathbf{S}_0\cdot \mathbf{S}_i\rangle(P)$ to yield the same value of $E(P)/T_N(P)$ for $P$ = 1~GPa. Because $dJ/dr$ was determined for exchange interactions along the $c$ axis, we ignore $J_2$ which is orthogonal to the $c$ axis and of low magnitude, and include only $J_1$ and $J_3$ in our calculation. From this rough calculation, we find that ordered moment $m = 2S\mu_B$ should decrease by 5\% at 1~GPa relative to ambient pressure, compared to the experimentally observed reduction of 6\% at 250~K or 15\% at 0~K based on extrapolation of the power-law fits in Fig.~\ref{fig:MagneticMoments}. We take the self-consistency of these results as confirmation of our proposed explanation for the pressure response of MnTe. 

From a theoretical viewpoint, these effects can be understood within the framework of the tight-binding model using the result that  $|J| = 4t^2/U$, where $J$ is the magnetic exchange interaction, $t$ is the hopping parameter of the tight-binding model, and $U$ is the on-site Coulomb repulsion~\cite{ashcroftmermin}. An increase in pressure induces orbital hybridization and leads to larger $t$, which in turn leads to larger $J$ and therefore a higher \TN\ (assuming $U$ does not change appreciably). This effect also makes the valence electrons more itinerant, which reduces the average local moment on each Mn site~\cite{bossini2020exchange, hafez2021magnetic}, consistent with our observations.  

\begin{figure*}
        \centerline{\includegraphics[width=1.0\textwidth]{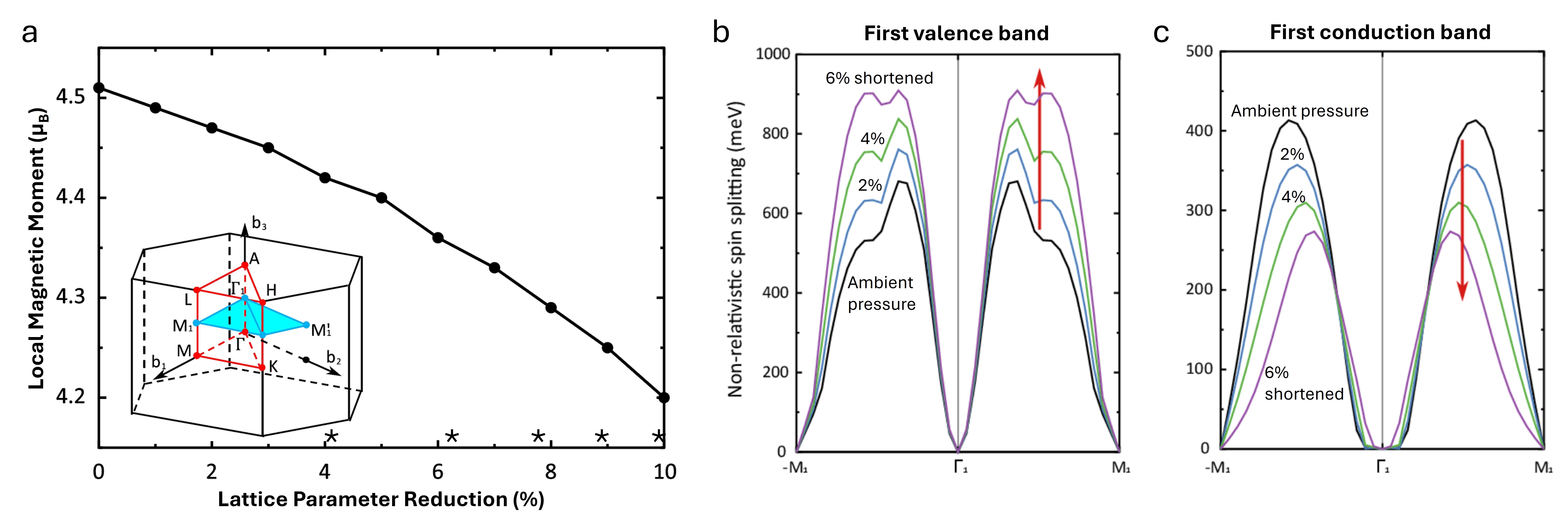}}
        \figcaption{\textbf{DFT-calculated response of the local magnetic moment and NRSS to increasing pressure.} (a) The magnitude of local magnetic moments calculated via DFT decrease steadily as the lattice parameters are reduced, simulating the experimental effect of pressure. The reference lattice parameters are taken from the DFT-relaxed structure with $a=b=4.22$~\AA, $c$=6.73~\AA. The asterisks just above the lower horizontal axis lattice paramaeter reductions corresponding to pressure values of 10, 20, 30, 40, and 50~GPa, calculated using the bulk modulus and equation of state given in Ref.~\cite{mimasaka1987pressure}. Inset: MnTe Brillouin zone. The blue region represents the $k_z=0.25(2\pi/c)$ plane ($M_1$-$\Gamma_1$-$M_1^{\prime}$ plane) where the NRSS is calculated. (b) The magnitude of the spin splitting in the first valence band in the $M_1$-$\Gamma_1$-$M_1^{\prime}$ plane for different lattice compressions, calculated as the absolute difference in energy between the spin-up and spin-down bands. The magnitude of the spin-splitting increases with pressure, as shown by the red arrow. (c) Same as (b) but for the first conduction band. Here, the spin-splitting magnitude decreases with pressure.}
        \label{fig:DFT}
    \end{figure*}

We also performed DFT calculations to verify this explanation from first principles. Using the experimentally observed lattice parameters at 250~K under ambient pressure and 1~GPa applied pressure, the calculations predicted the local moment to decrease from 4.48 to 4.47~\muB. Although quantitative accuracy is lacking, this result nevertheless provides qualitative agreement with our proposed explanation for the reduced local moment. To explore this trend further, we performed additional calculations in which we exaggerated the effect of pressure in our calculations by uniformly reducing the lattice parameters up to 10\% relative to those of the DFT-relaxed structure ($a=b=4.22$~\AA, $c=6.73$~\AA). The calculated local magnetic moment from these simulations is displayed as a function of lattice parameter reduction (a proxy for pressure) in Fig.~\ref{fig:DFT}(a), showing a significant drop from 4.51~$\mu_B$ at ambient pressure to 4.20~$\mu_B$ for 10\% reduction of the lattice parameters.
We performed the calculations both with and without SOC and found indistinguishable local moment magnitudes in both cases. We note that these results agree with other recent DFT calculations and x-ray emission spectroscopy data~\cite{wang;ChemMater22}. Another recent DFT study also predicts an increase in the exchange interactions with pressure~\cite{alaei2025}, again consistent with our explanation of the neutron diffraction results. These results do not depend sensitively on the chosen value of the on-site potential $U$ used in the DFT calculations, as evidenced by qualitatively similar results reported in Ref.~\cite{devaraj2024interplay} for $U = 3$~eV compared to the 4~eV used here.

It is also valuable to investigate the possibility of pressure-induced changes to the spin splitting in MnTe. From the DFT calculations, we extracted the spin-resolved band structure and confirmed spin-splitting with no SOC at low-symmetry $k$ points in the Brillouin zone (BZ). Detailed results are presented in the Supplemental Information~\cite{suppInfo}, including a discussion of symmetry elements that enforce the spin degeneracy in certain regions of the BZ and allow spin splitting in others. Here, we focus on the spin splitting observed in the $k_z=0.25(2\pi/c)$ plane ($M_1$-$\Gamma_1$-$M_1^{\prime}$ plane) as illustrated by the inset in Fig.~\ref{fig:DFT}(a). The magnitude of the spin splitting for different lattice compressions is shown in Fig.~\ref{fig:DFT}(b,c) for the first valence band (VB) and first conduction band (CB), respectively. The spin splitting in each band responds dramatically to pressure, changing by hundreds of meV for 6\% lattice compression. Interestingly, the VB and CB show the opposite trend with pressure: As the pressure increases, the spin splitting in the VB increases, whereas it decreases in the CB. This highlights the rich tunability of the NRSS attributes of MnTe with pressure and is consistent with previous DFT calculations~\cite{devaraj2024interplay}.


The dependence of the magnetic moment and \TN\ on pressure is likely to have interesting applications. For example, chemical pressure induced by replacing Te with the smaller chalcogenides Se or S may mimic the effect of physical pressure and raise \TN\ higher above room temperature, thereby expanding the temperature range available for practical applications while retaining other properties of MnTe. Indeed, an earlier report~\cite{chehab1986magnetic} indicates that \TN\ rises as expected when up to 20\% of the Te is replaced with Se. It may also be possible to induce a negative chemical pressure to lower \TN\ or increase the low-temperature ordered moment. The distinct VB- and CB-spin-splitting behavior with pressure likewise invites future experimental study and may enable novel applications in spintronics. Additional investigations will be necessary to explore these possibilities fully. Overall, this study provides a promising outlook on the use of pressure to optimize the properties of MnTe and potentially other NRSS antiferromganets.

\textbf{Data Availability}
The data that support the findings of this article are openly available~\cite{carlisle2025neutron}.

\textbf{Acknowledgements} We thank Malcolm Cochran of Oak Ridge National Laboratory for valuable assistance selecting the best pressure medium for the experiment, as well as Xiuwen Zhang for valuable discussions. The neutron scattering work carried out by EPC and BAF was supported by the U.S. Department of Energy, Office of Science, Basic Energy Sciences (DOE-BES) through Award No. DE-SC0021134. The DFT calculations of electronic properties performed by AZ and J-XX were supported by DOE-BES Materials Sciences and Engineering Division under Award No. DE-SC0010467. The DFT calculations used resources of the National Energy Research Scientific Computing Center, which is supported by the Office of Science of the U.S. Department of Energy. J-XX used the VASP software under the license from Alex Zunger. A portion of this research used resources at the High Flux Isotope Reactor, a DOE Office of Science User Facility operated by the Oak Ridge National Laboratory. The beam time was allocated to HB-2A on proposal number IPTS-32506. Sample synthesis and characterization at ORNL was supported by the U.S. Department of Energy, Office of Science, Basic Energy Sciences, Materials Sciences and Engineering Division.

\textbf{Author Contributions} BAF and RPH conceived of the initial project, with BAF providing overall supervision. EPC, GY, SC, and BH conducted the neutron diffraction experiment. EPC analyzed the neutron diffraction data, with help from GY. MAM provided the sample of MnTe, and MAM, RPH, and GY performed the sample characterization. AZ and J-XX performed the density functional theory calculations and provided input on the theory component. EPC and BAF wrote the manuscript, with input on the technical details of the density functional theory calculations and the introduction from AZ and J-XX. All authors read and approved the final manuscript.



\textbf{Competing Interests}
All authors declare no financial or non-financial competing interests.

\bibliography{pressure_abbrev}

\end{document}


\title{
	Supplementary Information: Tuning the magnetic properties of the spin-split antiferromagnet MnTe through pressure
}
	
\author{Edison P. Carlisle}
\affiliation{ %
    Department of Physics and Astronomy, Brigham Young University, Provo, Utah 84602, USA.}

\author{George Yumnam}
\affiliation{Materials Science and Technology Division, Oak Ridge National Laboratory, Oak Ridge, Tennessee 37830, USA}

\author{Stuart Calder}
\affiliation{Neutron Scattering Division, Oak Ridge National Laboratory, Oak Ridge, Tennessee 37831, USA}

\author{Bianca Haberl}
\affiliation{Neutron Scattering Division, Oak Ridge National Laboratory, Oak Ridge, Tennessee 37831, USA}

\author{Jia-Xin Xiong}
\affiliation{Renewable and Sustainable Energy Institute, University of Colorado Boulder, Boulder, Colorado 80309, USA}

\author{Michael A. McGuire}
\affiliation{Materials Science and Technology Division, Oak Ridge National Laboratory, Oak Ridge, Tennessee 37830, USA}

\author{Alex Zunger}
\affiliation{Renewable and Sustainable Energy Institute, University of Colorado Boulder, Boulder, Colorado 80309, USA}

\author{Rapha\"el P. Hermann}
\affiliation{Materials Science and Technology Division, Oak Ridge National Laboratory, Oak Ridge, Tennessee 37830, USA}

\author{Benjamin A. Frandsen}
\affiliation{ %
    Department of Physics and Astronomy, Brigham Young University, Provo, Utah 84602, USA.}
\email{benfrandsen@byu.edu}

\maketitle

\subsection{DFT calculations: Local magnetic moments and NRSS under pressure}

\begin{figure}[h!]
        \centerline{\includegraphics[width=0.5\textwidth]{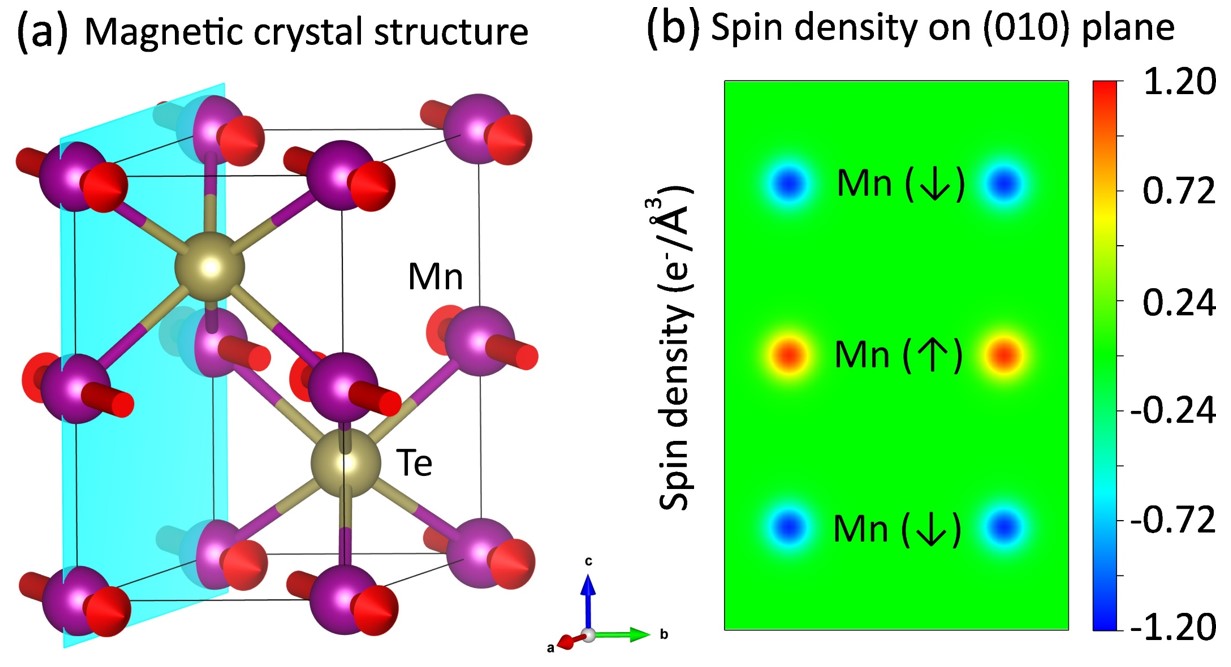}}
        \figcaption{(a) Collinear magnetic configuration of MnTe with local magnetic moments parallel or antiparallel to the [110] direction. (b)  DFT-calculated spin density in the (010) plane, illustrated by the blue plane in (a).}
        \label{fig:magstruc}
    \end{figure}

\begin{figure}[h!]
        \centerline{\includegraphics[width=0.5\textwidth]{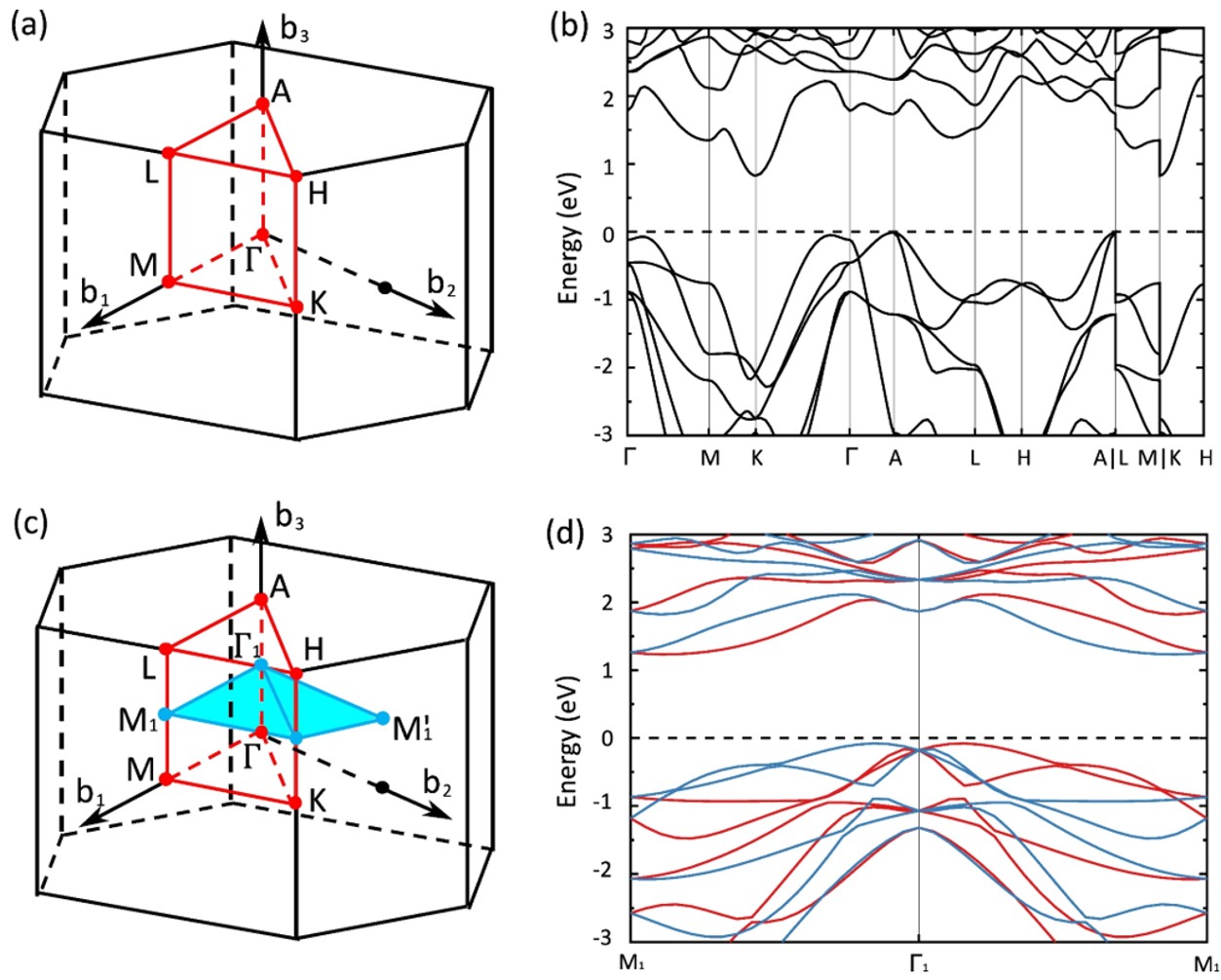}}
        \figcaption{\textbf{Band structure (SOC = 0) of MnTe at different reciprocal $k$-point lines by DFT calculations.}(a, b) No spin splitting at high-symmetric k-point lines. (c, d) Non-relativistic spin splitting at low-symmetric k-point lines. The Brillouin zone in (a, c) gives the high-symmetric and low-symmetric $k$-point lines marked in red and blue. The band structures in (b, d) show the energy dispersions along k-point lines in (a, c). The blue plane in (c) represents the $k_z=0.25(2\pi/c)$ plane ($M_1$-$\Gamma_1$-$M_1^{\prime}$ plane). The red and blue curves in (d) band structure represent opposite spin-polarized energy bands.}
        \label{fig:bandstruc}
    \end{figure}

    To complement the experimental results and provide a theoretical basis for the interpretation of the results, we performed density functional theory (DFT) calculations for MnTe under pressure. As seen in Fig.~\ref{fig:magstruc}(a), the collinear local magnetic moments on Mn atoms align parallel or antiparallel to the [110] direction within each Mn layer, stacked in an alternating A-type sequence along the $c$ axis.
    The DFT-calculated band structure with SOC set to zero is shown in Fig.~\ref{fig:bandstruc}.
    MnTe is an insulator with an indirect band gap with its conduction band minimum (CBM) and valence band maximum (VBM) at the $K$ and $A$ points, respectively [Fig.~\ref{fig:bandstruc}(b)]. There is no NRSS at high-symmetric $k$-points including $k_z$=0 plane ($M$-$\Gamma$-$K$ plane), $k_z$=0.5 plane ($L$-$A$-$H$ plane), and $k_z$ direction ($\Gamma$-$A$ line). The spin-degenerate energy bands on $k_z$=0 and $k_z$=0.5 plane are enforced by the (001) reflection-mirror plane of spin-opposite sublattices in the magnetic crystal structure [Fig.~\ref{fig:magstruc}(a)]. The spin degeneracy of the bands in the $k_z$ direction is protected by the $z$-direction six-fold screw axis of spin-opposite sublattices in the magnetic crystal structure [Fig.~\ref{fig:magstruc}(a)]. The NRSS in MnTe can only appear in lower-symmetry $k$-points, such as $k_z=0.25(2\pi/c)$ plane ($M_1$-$\Gamma_1$-$M_1^{\prime}$ plane) as shown in Fig.~\ref{fig:bandstruc}(d).

\bibliography{pressure}